\documentclass{article}
\usepackage{spconf,amsmath,graphicx}

\usepackage{hyperref}

\usepackage{mathrsfs,multicol,multirow,cite}
\usepackage{color}
\usepackage{url}

\title{A comparative study of generative models for child voice conversion}
\name{Protima Nomo Sudro, Anton Ragni, Thomas Hain}
\address{Department of Computer Science, The University of Sheffield}
\begin{document}
%
\maketitle

\begin{abstract}
Generative models are a popular choice for adult-to-adult voice conversion (VC) because of their efficient way of modelling unlabelled data.  To this point their usefulness in producing children speech and in particular adult to child VC has not been investigated. For adult to child VC, four generative models are compared: diffusion model, flow based model, variational autoencoders, and generative adversarial network. Results show that although converted speech outputs produce by those models appear plausible, they exhibit insufficient similarity with the target speaker characteristics. We introduce an efficient frequency warping technique that can be applied to the output of models, and which shows significant reduction of the mismatch between adult and child. The output of all the models are evaluated using both objective and subjective measures. In particular we compare specific speaker pairing using a unique corpus collected for dubbing of children speech.
\end{abstract}
\noindent\textbf{Index Terms}: Generative models, child speech, voice conversion, post processing

\section{Introduction}
\label{sec:intro}
  
Voice conversion (VC) aims to impersonate target speaker characteristics given source speech. In VC, the linguistic content, speaking style, pitch, intonation patterns, prosody are maintained, however, the speaker characteristics are varied~\cite{stylianou1998continuous}. VC techniques can be broadly classified as parallel and non-parallel methods based on the data used for training~\cite{sisman2020overview}. Parallel VC contains same utterances from source and target speakers and vice versa for non-parallel VC. Parallel VC such as Gaussian mixture model, non-negative matrix factorization, produce high quality converted speech, but it is challenging to acquire sufficient quantities of parallel speech data.  To overcome this issue, various non-parallel VC techniques have been proposed using a variety of different approaches such as generative models and recognition-synthesis based VC~\cite{vcc2020summary}. Generative model based VC learns the mapping function between source and target speech from non-parallel data using adversarial learning. The recognition-synthesis VC framework consists of two modules, the recognizer (or encoder) extracts the linguistic content and the synthesizer (or decoder) generates converted speech. 

VC has numerous applications including, customizing audiobook and avatar voices, computer assisted pronunciation training, speech to singing conversion, speech synthesis, communication aids for speakers with impaired speech, speech enhancement, voice assistants, talking robots, and dubbing. Dubbing is of particular interest for the current work, but it has limited literature~\cite{sisman2019singan}.

This study explores whether VC techniques can benefit dubbing the adult to child VC scenario in particular. The process of dubbing involves translating original dialogue from media based on the script, tone, genre, emotions and synchronizing them with lip movements. Specifically with children, dubbing is difficult because of limited children voice resources, expression of desired tone and mood during recording. To train a model for an adult to child VC system required for dubbing, we require acted speech data from professional voice talents. The acted speech data, however, is very different compared to read speech data used in typical VC~\cite{busso2008iemocap,du2021expressive}. 

Recent aproach to VC~\cite{popov2021diffusion} suggests the recognition-synthesis based VC as a state-of-the-art approach~\cite{liu2021any}. However,this approach requires large amounts of data for training. In low resource scenario like adult to child movie dubbing, large amount of data is hard to collect. On the other hand, generative model based VC were studied by researchers, where a smaller training dataset is used to train the VC system. 
In this paper four different generative modeling techniques that have shown promising results for adult speech VC are studied. These are generative adversarial networks (GAN), variational autoencoders (VAE), diffusion models, and flow based models. 

Furthermore, a post-processing system is introduced with the aim to compensate for a possible lack of similarity between converted and target speech. Several works in speech synthesis and speech enhancement have also reported improved performance when post-processing is applied for further refinement of predicted acoustic features~\cite{xu2017multi, wang2017tacotron}. In particular, the use of frequency warping is investigated in the current work, which has been previously used for mapping source to target speech characteristics in VC, speech synthesis, and ASR studies~\cite{erro2009voice, valentini2015towards,yeung2019frequency}.

In literature, only few studies have been reported on adult to child VC and even fewer on dubbing applications. One of the studies reported using read speech for training adult to child CycleGAN VC model for automatic speech recognition (ASR) application~\cite{shahnawazuddin2020voice}. The study was performed using PF-STAR corpus which consists of few read speech utterances. Other studies  by~\cite{mukhneri2020voice,watts2009synthesis} reported using Gaussian mixture model (GMM) based adult to child VC for speaker adaptation and dubbing using their own internal dataset. Our work is different from all of these in several important aspects: (a) first, none of the prior works considered using real media data from professional voice talents, (b) next, in~\cite{mukhneri2020voice}, dubbing was performed for Indonesian language using several words only,  (c) in\cite{shahnawazuddin2020voice}, VC was performed for data augmentation and did not investigate target child speaker similarity, (d) and finally, in ~\cite{watts2009synthesis} VC was applied to the output of a speech synthesiser.


Although, the previous studies showed effective performance for specific applications, mostly all the studies were carried out using read speech. Acted speech from an adult poses challenges to translate their expression and intonation to a child speech~\cite{turunen2017interplay}. Motivated by the limitations of the previous work adult to child speech dubbing, the contributions of this work are: 1) investigate four generative models for adult to child speech dubbing using acted speech data from professional and amateur voice talents.  2) introduce warping as a post-processing step for all the generative models, 3) the proposed method achieves improved results compared to standalone VC. To the best of our knowledge, this is the first study for adult-to-child speech dubbing using acted speech data.

The remaining paper is organized as follows. The generative models and post processing method are discussed in Section~\ref{sec2} and Section~\ref{sec3} respectively. The experimental details, results and discussions are reported in Section~\ref{sec4}. Finally, the work is concluded in Section~\ref{sec5}.

\section{Investigation of voice conversion models}
\label{sec2}

The generative models with its relaxed requirements towards training data offers a potential solution for adult to child VC.  Hence, we compare different class of generative models for child VC. This section briefly explains the VC workflow for each of the models in combination with warping.

\subsection{Generative adversarial network}
\begin{figure}[!t]
	{\centering
		{\includegraphics[width=0.48\textwidth]{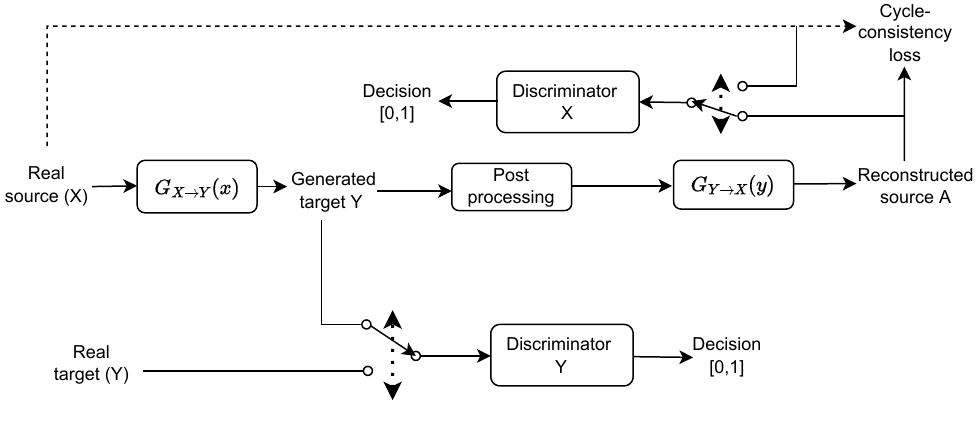} }
		\caption{Inference stage of the CycleGAN based VC system}
		\label{cycleganvc}
	}
\end{figure}
Generative adversarial networks (GAN) is a generative model that learns to map samples from one distribution to another distribution using non-parallel data~\cite{goodfellow2020generative}. Among various types of GANs, CycleGAN is considered to be one of the state-of-the-art techniques for adult-to-adult VC. Therefore, the CycleGAN-VC2 model is explored in this work for adult to child VC. Figure~\ref{cycleganvc} shows the framework for adult to child speech conversion using the CycleGAN-VC2 system combined with warping. 

The CycleGAN-VC2 architecture is implemented based on the approach reported in~\cite{kaneko2019cyclegan}. The CycleGAN-VC2 model is tasked to learn forward and inverse transformations simultaneously to find an optimal source and target pair from non-parallel data using adversarial and cycle-consistency loss. Basically, the generator $G_{X\rightarrow Y} (x)$ maps the source distribution $x$ into target distribution $y$. On the other hand, the generator $G_{Y\rightarrow X} (y)$ maps the generated target distribution into source distribution. Hence, the  generator $G_{Y\rightarrow X} (y)$ enforces the sequence of transformation is similar to original input. During inference, the converted features from the generator $G_{X\rightarrow Y} (x)$ is input to the frequency warping module, which is finally passed through WORLD vocoder to obtain converted speech.  

CycleGAN-VC2 performs effectively in a single source-target speaker pair. However, its behaviour for multi-speaker scenario and out of sample generalization setting is not known.



\subsection{ Variational autoencoder}	
Variational autoencoder(VAE) refers to a latent based model which learns complex data distributions in a generative way with prior and posterior distributions. The  latent based model is optimized by learning the joint distribution of the conditional data distribution and uniform
\begin{figure}[!b]
	{\centering
		{\includegraphics[width=0.48\textwidth]{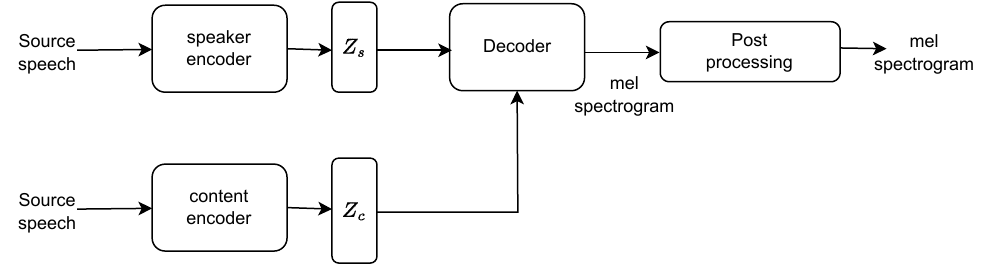} }
		\caption{Inference stage of the VAE based VC system }
		\label{vaevc}
	}
	\end{figure}
 distribution of the latent variable~\cite{chou2019one}. The VAE based VC system is shown in Figure~\ref{vaevc}, where post-processing is applied on the output of the decoder.  It consists of a speaker encoder, a content encoder and a decoder. The speaker encoder is trained to generate speaker representations $Z_s$ and content encoder is trained to generate content representation $Z_c$. The objective function of the VAE training is to minimize the combination of reconstruction loss and Kullback Leibler (KL) divergence loss with weighted hyperparameters. The KL divergence is used to match the posterior distribution with the prior distribution. 


Compared to CycleGAN, VAE can generalize for multi-speaker settings and unseen speakers in the training stage. However, the output quality is affected due to the combined optimization of reconstruction loss and KL divergence loss.

\subsection{ Flow based generative model}	
The flow based generative model builds complex distribution from simple distribution via a flow of successive(invertible) transformations~\cite{serra2019blow}. Flow based models present a constructive way to estimate the data distribution, which allow exact inference and likelihood evaluation, and exact posterior inference.

\begin{figure}[!t]
	{\centering
		{\includegraphics[width=0.48\textwidth]{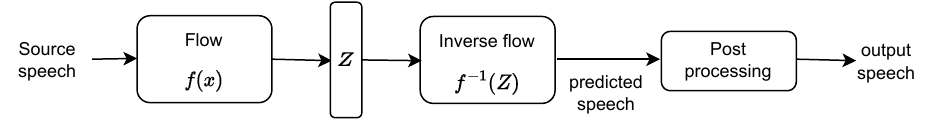} }
		\caption{Flow based VC system}	
		\label{blowvc}
	}
	\end{figure}
The flow based model approach followed in this work learns a bijective mapping function from input $x \in \mathcal{X}$ to latent representations $z \in \mathcal{Z}$, where, $z = f(x)$ and $x =f^{-1}(z)$. The mapping function $f$ is an invertible function which is regarded as a normalizing flow. The mapping function $f$ consists of a sequence of invertible transformations. The transformations are learned via maximum likelihood objective function. In the flow based VC system depicted in Figure.~\ref{blowvc}, the model is trained to convert speech in an end-to-end manner using raw audio on a frame basis without context. Each of the frame has a speaker identifier labels. The flow based VC use source speaker identifier for transforming $x$ to $z$ and a target speaker identifier for transforming $z$ to converted speech frame.

Flow based model differs from CycleGAN-VC2 and VAE in terms of computing log-likelihood exactly. In case of VAE only lower bound on the log-likelihood can be approximated and in CycleGAN-VC2, log-likelihood estimation is avoided and instead optimize min-max objective. In addition, flow based model allows exact posterior inference using invertible function. Whereas, VAE parameterize approximate posterior and in CycleGAN-VC2 there is no latent variable inference.

\subsection{ Diffusion model}	
Diffusion probabilistic models destroy the original structure of a data sample using some fixed process and reconstruct the data from the corrupted noisy distribution by reversing the process in time. Basically, diffusion implies transforming a complex (any data) distribution  into a simple (predefined) prior distribution (e.g., Gaussian) over long time. 
\begin{figure}[!b]
	{\centering
		{\includegraphics[width=0.48\textwidth]{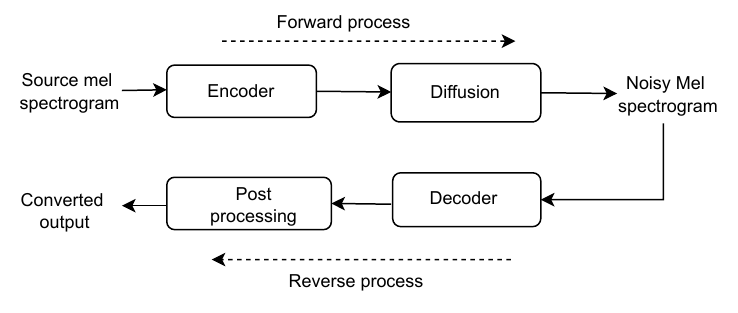} }
		\caption{Diffusion model based VC system}	
		\label{diffvc}
	}
	\end{figure}
A block diagram of the diffusion model based VC system is shown in Figure.~\ref{diffvc}. A diffusion probabilistic models (DPM) comprises of two process: forward diffusion and reverse diffusion. The forward process gradually destroy data by adding a noise schedule over a number of steps, which results in approximate normal distribution. While the reverse diffusion is trained to remove noise. The current work follows the diffusion model based VC reported in~\cite{popov2021diffusion}, where VC is accomplished by considering the forward diffusion process equivalent to an encoder and the reverse diffusion process is regarded as a decoder model. The encoder is trained to obtain speaker independent representations and is implemented using the Transformer based architecture. The reverse diffusion is defined in terms of a stochastic process based on Ito calculus. The DPM is trained to minimize the distance (reconstruction error) between forward and reverse process. The reverse process is conditioned on the target speech characteristics to provide information about the target speaker.

\section{Post-processing}
\label{sec3}
Post-processing is usually applied to improve converted speech quality in speech synthesis and speech enhancement studies~\cite{demiroglu2017postprocessing,ramakrishnan2011efficient}. Post-processing is also applied in speech recognition studies to improve ASR performances~\cite{sodhi2021mondegreen}. Motivated by these studies, in our current work, warping is applied  as a post-processing method to improve the converted speech similarity with the target speaker should any of the generative models exhibit weak inherent VC performance.

Warping is a technique commonly used to match the vocal tract length characteristics between some source and some target speaker. It has been used in automatic speech recognition~\cite{yeung2019frequency}, speech synthesis and VC~\cite{valentini2015towards,erro2009voice}.  In this study, warping is explored to address possible mismatch between real child speech and converted speech. Following \cite{valentini2015towards}, the warping of frequency spectrum can be achieved by multiplying mel cepstral coefficients ${c}=c_{1},\ldots,c_{M}$ with the learnt warped matrix ${D}=\{D_{k,m}\}$,  The learning starts by initialising
\begin{eqnarray}\label{warp}
{D}_{k,m} = cos(m \tilde{\omega}_{k,m})
\end{eqnarray}
where, $k = 0,1,2,..., N-1$ denotes the frequency scale of fast Fourier transform and $m=0,1,2,..., M-1$ denotes the mel frequency scale. The phase characteristics, $\tilde{\omega}_{k,m}$ in equation~\ref{warp} is given by,

\begin{eqnarray}
\tilde{\omega}_{k,m} &=& tan^{-1} \left(\frac{(1-\alpha^2 )sin\left(2\pi \frac{f_{m}}{f_s}k\right)}{(1+\alpha^2) cos\left(2\pi\frac{f_{m}}{f_s}k\right) -2\alpha}\right) \\
\alpha &=& \frac{sin (\pi (f_{src} - f_{tgt} )/f_s)}{cos (\pi (f_{src} + f_{tgt} )/f_s)}
\end{eqnarray}
here, $\alpha$ is the warping factor, $f_{src} ,f_{tgt}, f_s$, and $f_m$ denotes source frequency, target frequency, sampling frequency the mel frequency scale respectively.
In order to learn warping by training a network, the cost function of the warped spectrum is given by, 
\begin{eqnarray}
E({D}) = \displaystyle \sum_{k=0}^{N-1} \left( \sum_{m=0}^{M-1} D_{k,m} c_m - \sum_{m=0}^{M-1}D^{*}_{k,m} c_m^*\right)^2
\end{eqnarray}
$c_m$ and $c_m^*$ corresponds to the $m^{th}$ order mel cepstral coefficients of source speech and reference target speech respectively. $D^{*}_{k,m}$ is an $N \times M$ matrix that corresponds to the reference spectrum given by,
\begin{equation}
D_{k,m}^{*} = cos\left(m 2\pi \frac{f_m}{f_s} k\right)
\end{equation}

The learnt ${D}$ is used to obtain the warped frequency scale and obtain the improved features. 


\section{Experiments}
\label{sec4}
This section demonstrates the experiments performed for the generative model based VC with and without warping. 
\subsection{Corpus}
To study the performance of adult to child VC for the aforementioned generative models, an internal dataset recorded in a studio by professional dubbing actors is used. The dataset contains 5 hours of audio from 3 speakers (adult male, adult female, and female child). The acted speech data consists of utterances recorded from several child centered movies. The audio is resampled at 16 kHz and randomly extract $10\%$ of the data for validation and $10\%$ for testing. 

\subsection{Implementation details}


CycleGAN-VC2 based VC is implemented as reported in~\cite{kaneko2019cyclegan} using 36 dimensional Mel cepstral coefficients extracted using WORLD analyzer~\cite{morise2016world}. 
The initial learning rate of the generator and discriminators are 0.0002 and 0.0001, respectively. In case of VAE based VC, the framework described in~\cite{chou2019one} is used, where inputs are 512 dimensional Mel scale spectrograms and Griffin-Lim algorithm is used to reconstruct the wave files for a window length of 50 ms and 12.5 ms hop size. The learning rate of the optimizer is 0.0005. For the flow based VC, the system described in~\cite{serra2019blow} is used. The training is done on raw waveform data. The training is initialized with the learning rate of 0.0001. The system is analyzed using 256 ms window length with no overlap and Griffin-Lim algorithm is used to reconstruct the wave files. For diffusion model the system shown in~\cite{popov2021diffusion} is followed. Here, 80-dimensional Mel-spectrograms are used as input features and the noise schedule is a non-negative function, whose parameters are set to 0.05 and 20.0. HiFi-GAN vocoder is used to reconstruct the wave files for the converted speech. The speech is analyzed for a window of length 1024 and hop size of 256.

All the generative models are first studied without the post processing. Next study reports the VC along with post processing. The warping function is trained using 80-dimensional Mel features. Mean squared error is used to minimize the difference between converted features and the reference target features as described in Section~\ref{sec3}.

\subsection{Evaluation metrics}
To assess the converted speech output, Mel-cepstrum distortion (MCD)~\cite{kubichek1993mel} and normalized F0 root mean square error (F0 RMSE) are used as the evaluation metrics. MCD computes the Euclidean distance between two vectors that describes the spectral characteristics. The MCD measures spectral distortion between adult and child Mel cepstral coefficients and is obtained by computing the mean of MCD values across all the frames. Similarly, the F0 RMSE measures the F0 difference between converted and target speech and is computed by averaging across all the values.

\subsection{Results and discussion}
\begin{table}[tbh!]
	\centering
	\caption{MCD and F0 RMSE values for different generative model based VC approaches. * here denotes VC combined with warping }
	\label{t1}
\scalebox{0.9}{		
\begin{tabular}{ccccc}
\hline \hline
Methods          & MCD             & MCD*   & F0 RMSE    & F0 RMSE* \\ \hline
CycleGAN         & 8.47                &8.29       & 0.67               & 0.55       \\
VAE              & 9.02                &8.90       & 0.70               & 0.62\\ 
Flow             & 9.67               &9.52       & 0.42                 & 0.36\\ 
Diffusion        & 7.82               &7.47       & 0.39                  & 0.31\\ \hline\hline
\end{tabular}
}
\end{table}

\begin{table}[t!]
	\centering
	\caption{Performance of different generative model based VC approaches for CMU multi-speaker child data. * here denotes VC combined with warping}
	\label{t3}
\scalebox{0.9}{		
\begin{tabular}{ccccc}
\hline \hline 
Methods              & MCD       & MCD*               & F0 RMSE     & F0 RMSE*      \\ \hline        
CycleGAN             & 8.08      &7.98               & 0.63            & 0.52        \\  
VAE                  & 9.75       & 9.61              & 0.81           &0.67           \\    
Flow                 & 9.92      &9.18               & 0.79           & 0.70           \\ 
Diffusion            & 8.92     &8.45               & 0.72           & 0.68\\ \hline  \hline    
\end{tabular}
}
\end{table}

The results are summarized in Table~\ref{t1}, Table~\ref{t3} for the internal acted speech data and CMU corpus respectively. From both the tables, it can be observed that CycleGAN and Diffusion model showed improved performance compared to VAE and Flow based models. 

When warping is performed along with VC, it is observed that both MCD and F0 RMSE values have decreased for all the models. It is to be noted from Table~\ref{t1} that the results are obtained for single speaker pair. In addition to internal acted speech data, experiments are carried out using multi-speaker child corpus. For this purpose, publicly available CMU speech corpora is used~\cite{eskenazi1997cmu}. The CMU child speech corpora contains 5191 utterances (9.1 hour) from 76 speakers, where there are 52 female and 24 male speakers respectively. The vocabulary size is 876. The age of the child speakers ranges from 6-11 years. As there is no parallel adult speech data for the CMU kids corpora, we recorded all the utterances of CMU kids corpora by an adult dubbing actor. Then VC is performed with and without warping between adult actress and child speakers from CMU kids corpus. The VC results for CMU kids corpus is shown in Table~\ref{t3}.

From Table~\ref{t3}, consistent results are observed similar to table~\ref{t1}. This implies that generative models without warping can be used to perform VC. However, with warping, the speaker similarity is further increased. Next part analyze the studied approach for different speaker pair combinations. The respective results are shown in Figure~\ref{spk_pair}.

\begin{figure}[!tbh]
	{\centering
  		{\includegraphics[width=0.4\textwidth,height=2.5 cm, scale=9.5]{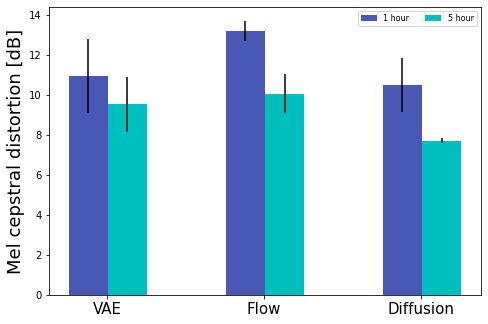} }
		\caption{Average MCD values for different amount of training data (1 hours and 5 hours) used }	
		\label{duration}
	}
	\end{figure}

From Table~\ref{spk_pair}, it appears that results vary when the difference between the speaker characteristics increases in terms of pitch, vocal tract characteristics and style. 
The behaviour for different amount of training data for VAE, Flow, and diffusion models is depicted in Figure~\ref{duration}. Since, 5 hours of parallel data of a single speaker pair is not available,  only 1 hour and 2 hour of training data for CycleGAN-VC2 are compared. With 1 hour and 2 hour training data for CycleGAN-VC2, mean an standard deviation values are: $9.07 \pm 0.65$ and $8.92 \pm 0.93$ respectively. With respect to different amount of training data used to perform VC, it is observed that Diffusion model and flow based model showed most significant reduction compared to CycleGAN and VAE. Though, Flow based VC showed significant MCD reduction with more amount of training data, its MCD value is comparatively highest among others. The reason may be attributed to the fact that, for an end-to-end raw speech synthesis, 1 and 5 hour of data is not sufficient. In case of CycleGAN-VC2, the improvement is minimal, the reason being that, the CycleGAN-VC2 architecture is implemented in a one-to-one speaker pair features. That is unlike other models, speaker representations and content representations are not disentangled. The low performance of VAE model may be caused by over-smoothing. Among the four generative models, diffusion model based VC showed signicant scaling with 5 hour of training data. Additionally, VC system using diffusion model exhibits lowest MCD values among all the 4 models. The improved quality of diffusion model based VC is due to the diffusion probabilistic model based decoder, which employs forward and reverse diffusion to generate data.

\begin{table}[!t]
\centering
	\caption{MCD values for different speaker pair combinations. C,F,M denotes child, female, and male}
	\label{spk_pair}
\scalebox{0.9}{		
\begin{tabular}{ccccc}
\hline\hline
\begin{tabular}[c]{@{}c@{}}Conversion\\ pair\end{tabular} & CycleGAN & VAE  & Flow  & Diffusion \\ \hline
F-C                                                       & 8.43     & 8.57 & 9.83  & 7.77      \\
M-C                                                       & 8.51     & 9.48 & 11.5 & 7.87      \\
F-M                                                       & 9.31     & 8.27 & 8.72  & 7.68      \\
M-F                                                       & 10.03    & 11.8 & 10.2  & 7.50       \\ \hline
Average                                                   & 9.07     & 9.53 & 10.06 & 7.71 \\ \hline\hline    
\end{tabular}
}
\end{table}

\begin{table}[tbh!]
\centering
 	\caption{Mean opinion scores of the reference and converted speech evaluated by native english listeners}
 	\label{mos}
\scalebox{0.8}
{	 	
\begin{tabular}{c|cc|cc}
\hline \hline
\multirow{2}{*}{Methods} & \multicolumn{2}{c|}{Naturalness} & \multicolumn{2}{c}{Similarity} \\ \cline{2-5} 
                         &  nowarp               & warp           & nowarp               & warp          \\ \hline
CycleGAN          & 3.34 $\pm$ 0.86   & 3.3 $\pm$ 0.81   & 3.2 $\pm$  0. 92 & 3.29 $\pm$ 0.64 \\
VAE               & 3.05 $\pm$ 0.95   & 3.02 $\pm$ 0.96  & 2.79 $\pm$ 0.77  & 2.88 $\pm$ 0.6  \\
Flow              & 2.99 $\pm$ 1.19   & 2.98 $\pm$ 1.1   & 2.65 $\pm$ 0.89  & 2.72 $\pm$ 0.85 \\
Diffusion         & 3.35 $\pm$ 0.98   & 3.33 $\pm$ 0.9   & 3.01 $\pm$ 0.95  & 3.13 $\pm$ 0.79 \\ \hline \hline
\end{tabular}
}
\end{table}

\subsection{Subjective evaluation}
In this subsection, subjective evaluation studies are presented. The study is conducted for both naturalness and similarity of the target speech with VC outputs (with and without warping). A total of 12 native english speakers participated in the study. Each of them were provided with 10 utterance pairs for each respective VC approach. The study was conducted in a quiet room using same headphone and laptop. All the tests were performed in a similar settings for all the speakers.

To evaluate the speech samples, the listeners were asked to rate the perceptual quality of the speech using a 5-point rating scale, where 1$=$bad, 2$=$poor, 3$=$fair, 4$=$good, and 5$=$excellent. The average scores across all the listeners are computed to obtain the mean opinion score (MOS) for different generative model based VC approaches. The subjective evaluation results for VC with warping and without warping are shown in Table~\ref{mos}. From the MOS values, it is observed that after warping the similarity is improved for all the models, and the naturalness is comparable to the VC output without warping. For demo some of the speech samples are made available in the given link 
\url{https://drive.google.com/drive/folders/17DnkZurpMTg7sMiM-N8TMw-nZQGEbEtD?usp=share_link}
\section{Conclusion}
\label{sec5}
In this work, a comparative analysis of four generative models for adult to child VC is presented. In order to improve the target speaker similarity, warping is performed as a post processing approach along with VC. The proposed approach is evaluated across different speech corpora and speaker pair combinations. From the evaluation results, it is observed that CycleGAN-VC2 showed improved results compared to other models. However, it performs only one-to-one mapping and need separate training for multiple speakers. On the other hand, VAE, flow and diffusion models can handle multiple speakers with no parallel training data and it is desirable for multi-speaker movie dubbing. When post-processing is applied, VAE, flow and diffusion models showed significant difference except CycleGAN-VC2. In the future work, VAE, flow and diffusion models will be explored to study the prosody and intonation patterns for child voice conversion.

\bibliographystyle{IEEEbib}
\bibliography{child_vc}

\end{document}